\documentclass[aps,prd,superscriptaddress,preprint,tightenlines,nofootinbib,showpacs]{revtex4}
\usepackage{amsmath,amssymb}
\usepackage{graphicx}
\usepackage{dcolumn}
\usepackage{bm}
\usepackage{epsfig}

\newcommand{\be}{\begin{equation}}
\newcommand{\ee}{\end{equation}}
\newcommand{\bea}{\begin{eqnarray}}
\newcommand{\eea}{\end{eqnarray}}
\newcommand{\etal}{{\it et al.}}

\begin{document}
%\preprint{CLEO Draft 08-28}
\preprint{CLNS 08/2046}
\preprint{CLEO 08-28}
\title{Search for {\boldmath {\boldmath $D^0\rightarrow \bar{p} e^+ $} and $D^0 \rightarrow p e^- $} }

\author{P.~Rubin}
\affiliation{George Mason University, Fairfax, Virginia 22030, USA}
\author{N.~Lowrey}
\author{S.~Mehrabyan}
\author{M.~Selen}
\author{J.~Wiss}
\affiliation{University of Illinois, Urbana-Champaign, Illinois 61801, USA}
\author{R.~E.~Mitchell}
\author{M.~R.~Shepherd}
\affiliation{Indiana University, Bloomington, Indiana 47405, USA }
\author{D.~Besson}
\affiliation{University of Kansas, Lawrence, Kansas 66045, USA}
\author{T.~K.~Pedlar}
\affiliation{Luther College, Decorah, Iowa 52101, USA}
\author{D.~Cronin-Hennessy}
\author{K.~Y.~Gao}
\author{J.~Hietala}
\author{Y.~Kubota}
\author{T.~Klein}
\author{R.~Poling}
\author{A.~W.~Scott}
\author{P.~Zweber}
\affiliation{University of Minnesota, Minneapolis, Minnesota 55455, USA}
\author{S.~Dobbs}
\author{Z.~Metreveli}
\author{K.~K.~Seth}
\author{B.~J.~Y.~Tan}
\author{A.~Tomaradze}
\affiliation{Northwestern University, Evanston, Illinois 60208, USA}
\author{J.~Libby}
\author{L.~Martin}
\author{A.~Powell}
\author{G.~Wilkinson}
\affiliation{University of Oxford, Oxford OX1 3RH, UK}
\author{H.~Mendez}
\affiliation{University of Puerto Rico, Mayaguez, Puerto Rico 00681}
\author{J.~Y.~Ge}
\author{D.~H.~Miller}
\author{I.~P.~J.~Shipsey}
\author{B.~Xin}
\affiliation{Purdue University, West Lafayette, Indiana 47907, USA}
\author{G.~S.~Adams}
\author{D.~Hu}
\author{B.~Moziak}
\author{J.~Napolitano}
\affiliation{Rensselaer Polytechnic Institute, Troy, New York 12180, USA}
\author{K.~M.~Ecklund}
\affiliation{Rice University, Houston, TX 77005, USA}
\author{Q.~He}
\author{J.~Insler}
\author{H.~Muramatsu}
\author{C.~S.~Park}
\author{E.~H.~Thorndike}
\author{F.~Yang}
\affiliation{University of Rochester, Rochester, New York 14627, USA}
\author{M.~Artuso}
\author{S.~Blusk}
\author{S.~Khalil}
\author{J.~Li}
\author{R.~Mountain}
\author{K.~Randrianarivony}
\author{N.~Sultana}
\author{T.~Skwarnicki}
\author{S.~Stone}
\author{J.~C.~Wang}
\author{L.~M.~Zhang}
\affiliation{Syracuse University, Syracuse, New York 13244, USA}
\author{G.~Bonvicini}
\author{D.~Cinabro}
\author{M.~Dubrovin}
\author{A.~Lincoln}
\author{M.~J.~Smith}
\author{P.~Zhou}
\author{J.~Zhu}
\affiliation{Wayne State University, Detroit, Michigan 48202, USA}
\author{P.~Naik}
\author{J.~Rademacker}
\affiliation{University of Bristol, Bristol BS8 1TL, UK}
\author{D.~M.~Asner}
\author{K.~W.~Edwards}
\author{J.~Reed}
\author{A.~N.~Robichaud}
\author{G.~Tatishvili}
\author{E.~J.~White}
\affiliation{Carleton University, Ottawa, Ontario, Canada K1S 5B6}
\author{R.~A.~Briere}
\author{H.~Vogel}
\affiliation{Carnegie Mellon University, Pittsburgh, Pennsylvania 15213, USA}
\author{P.~U.~E.~Onyisi}
\author{J.~L.~Rosner}
\affiliation{Enrico Fermi Institute, University of
Chicago, Chicago, Illinois 60637, USA}
\author{J.~P.~Alexander}
\author{D.~G.~Cassel}
\author{J.~E.~Duboscq\thanks{Deceased}}
\author{R.~Ehrlich}
\author{L.~Fields}
\author{L.~Gibbons}
\author{R.~Gray}
\author{S.~W.~Gray}
\author{D.~L.~Hartill}
\author{B.~K.~Heltsley}
\author{D.~Hertz}
\author{J.~M.~Hunt}
\author{J.~Kandaswamy}
\author{D.~L.~Kreinick}
\author{V.~E.~Kuznetsov}
\author{J.~Ledoux}
\author{H.~Mahlke-Kr\"uger}
\author{J.~R.~Patterson}
\author{D.~Peterson}
\author{D.~Riley}
\author{A.~Ryd}
\author{A.~J.~Sadoff}
\author{X.~Shi}
\author{S.~Stroiney}
\author{W.~M.~Sun}
\author{T.~Wilksen}
\affiliation{Cornell University, Ithaca, New York 14853, USA}
\author{J.~Yelton}
\affiliation{University of Florida, Gainesville, Florida 32611, USA}
\collaboration{CLEO Collaboration}
\noaffiliation

\date{April 9, 2009}

\begin{abstract}
We search for simultaneous baryon and lepton number violating decays of the $D^0$ meson. Specifically, we use 281 pb$^{-1}$ of data taken on the $\psi(3770)$ resonance with CLEO-c detector at the CESR collider to look for decays $D^0\rightarrow \bar{p} e^+$, $\bar{D}^0\rightarrow \bar{p} e^+$, $D^0\rightarrow p e^-$ and $\bar{D}^0\rightarrow p e^-$. We find no significant signals and set the following branching fraction upper limits: $D^0\rightarrow \bar{p} e^+ (\bar{D}^0\rightarrow \bar{p} e^+) < 1.1 \times 10^{-5} $ and $D^0\rightarrow p e^- (\bar{D}^0\rightarrow p e^-) < 1.0 \times 10^{-5} $, both at 90\% confidence level.
\end{abstract}

\maketitle
\section{Introduction}
Various Grand Unified Theories (GUTs) \cite{GUT} and many Standard Model (SM) extensions such as superstring models \cite{Lazarides} and supersymmetry (SUSY) \cite{SUSY} predict baryon number violation, and as a consequence nucleons can have finite, if long, lifetimes. However nucleon decay has not yet been observed \cite{PDG}. In all these theories baryon ($B$) and lepton ($L$) number violations are allowed but the difference  $\Delta (B-L) = 0$ is conserved. A higher generation SUSY model \cite{Nag} predicts decay modes having such $B$ and $L$ violating decays for $\tau$ leptons and for $D$ and $B$ mesons. The search for such $\tau$ decays has been performed \cite {CLEO, BELLE}, but decays of heavy quarks have not previously been investigated.

In this paper we describe a search for the $D$ meson decay channels $D^0\rightarrow \bar{p} e^+$, $\bar{D}^0\rightarrow \bar{p} e^+$, $D^0\rightarrow p e^-$ and $\bar{D}^0\rightarrow p e^-$. Such decays simultaneously violate $B$ and $L$ but conserve $\Delta (B-L)$. Several models of proton decay, e.g. in GUT, superstrings and SUSY as described above can be augmented to provide predictions on possible decay mechanisms.

\begin{figure}[htb]
\begin{center}
\centerline{ \epsfxsize=3.2in \epsffile{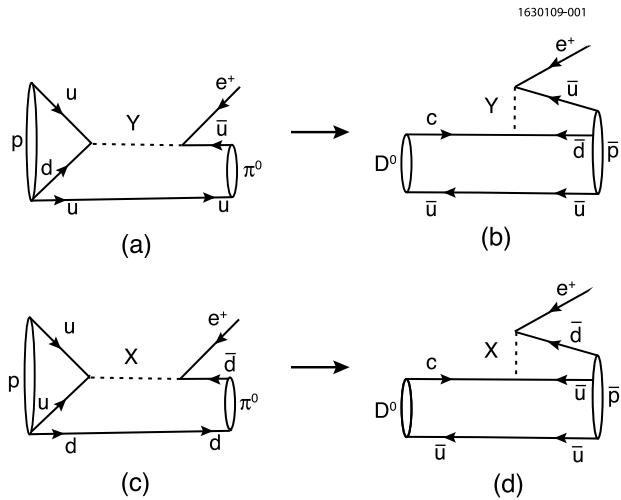}} \caption{
(a) and (c) are s-channel decay diagrams of $p \rightarrow \pi^0 e^+$ described by SU(5) theory and (b) and (d) are decay diagrams of $D^0 \rightarrow \bar{p} e^+$ based on analogous couplings.}\label{fig:GUT}
\end{center}
\vspace{-20pt}
\end{figure}

In SU(5) theory, protons can decay into several modes; one of them is $p\rightarrow e^+ \pi^{0}$. Biswal \etal~\cite{SU(5)} suggested five different decay diagrams. The decays are mediated by heavy hypothetical gauge bosons called $X$ and $Y$. The $X$ and $Y$ bosons have electric charge ${4 \over 3}e$ and ${1 \over 3}e$ and couple a quark to a lepton, hence they are sometimes called ``lepto-quarks." Figs.~\ref{fig:GUT}(a) and (c) show two of these possibilities that proceed via the $s$-channel. Fig.~\ref{fig:GUT}(b) is an analogous decay diagram for $D^0\rightarrow \bar{p} e^+$, where the mediator is a $Y$ boson. Here we take the coupling $e^+ Y \bar{u}$ as shown in Fig.~\ref{fig:GUT}(a) and introduce a coupling $c Y \bar{d}$ replacing a $u$ with a $c$ in the $t$-channel version of Fig.~\ref{fig:GUT}(a). Similarly, Fig.~\ref{fig:GUT}(d) shows another analogous decay diagram for $D^0\rightarrow \bar{p} e^+$ with an $X$ boson as the mediator; here we take the coupling of $e^+ X \bar{d}$ from Fig.~\ref{fig:GUT}(c) and use the coupling $c X \bar{u}$ by replacing a $u$ with a $c$ in the $t$-channel version of Fig.~\ref{fig:GUT}(c). The spectator in both decay diagrams is $\bar{u}$. No tree level diagrams allow $D^0\rightarrow p e^-$ in SU(5). However, a decay model can be constructed using higher order diagrams.
Arnowitt and Nath also predict proton decay in an $R$-parity violating \cite{R-parity} superstring based model that can also accommodate
$D^0\rightarrow \bar{p} e^+$ decay \cite{SuperString}.
\section{Data Sample, Signal Selection and Reconstruction Method}
We study the decays of $\overset{\scriptscriptstyle{(-)}}{D^0}\rightarrow \bar{p} e^+$ and $p e^-$ using the CLEO-c detector \cite{cleoc}. We do not assume that the two modes, $D^0 \rightarrow \bar{p} e^+$ and $D^0 \rightarrow p e^-$ are charge-parity (CP) conserved decays. When we refer to $\mathcal{B}(\overset{\scriptscriptstyle{(-)}}{D^0} \rightarrow \bar{p} e^+)$ we mean either $\mathcal{B}(D^0 \rightarrow \bar{p} e^+)$  or $\mathcal{B}(\bar{D}^0 \rightarrow \bar{p} e^+)$. Likewise, $\mathcal{B}(\overset{\scriptscriptstyle{(-)}}{D^0} \rightarrow p e^-)$ is shorthand for either $\mathcal{B}(D^0 \rightarrow p e^-)$  or $\mathcal{B}(\bar{D}^0 \rightarrow p e^-)$.

The CLEO-c detector consists of a CsI(Tl) electromagnetic calorimeter, an inner vertex drift chamber, a central drift chamber, and a ring imaging Cherenkov (RICH) detector inside a superconducting solenoid magnet providing a 1.0 T magnetic field. In this study we use 281 pb$^{-1}$ of CLEO-c data produced in $e^+$ $e^-$ collisions and recorded at the $\psi(3770)$ resonance. At this energy, the events consist of a mixture of $D^+ D^-$, $D^0 \bar{D}^0$ and $e^+e^- \rightarrow q\bar{q} (q=u, d, s)$ continuum events with a small number of $\tau^+ \tau^-$ and $\gamma \psi(2S)$ events.

We examine all the recorded events and look for $D^0$ candidates corresponding to $\overset{\scriptscriptstyle{(-)}}{D^0} \rightarrow \bar{p} e^+$ and $p e^-$. The selection criteria for charged tracks are similar to that described in \cite{bonvinci}, except that the momenta are required to be in the range from $50$ MeV/$c$ to 2 GeV/$c$. Moreover, we require that the polar angles, that the $p (\bar{p})$ and $e^-(e^+)$ subtend with respect to the beam axis are required to satisfy $|\cos\theta | \leq 0.9$. Protons are identified using only the energy loss information $(dE/dx)$ from the tracking chambers, since the kinematic limit of their momentum (900 MeV/$c$) is below threshold for radiation in the RICH detector. On the other hand, we do use the RICH, in combination with $dE/dx$, to aid in identification and elimination of kaons when the momentum is above 700 MeV/$c$, which is sufficiently above the RICH kaon radiation threshold. The specific requirements are discussed in Ref.~\cite{bonvinci}. Defining $\sigma_p$ as the difference between the expected ionization loss for a proton and the measured loss divided by the measurement error, with analogous definitions for $\pi$, $K$ and $e$, we require $|\sigma_p|<2.5$, $|\sigma_{\pi}|>3$, $|\sigma_K|>3$ and $\sigma_p^2 - \sigma_e^2<0$. We find that, for the momentum range of 0.5 to 0.9 GeV/$c$ the proton identification efficiency is 98\% and the probability that a pion (kaon) is misidentified as a proton is 0.9\% (1.6\%).

Electrons (positrons) are selected as in Ref.~\cite{elec}, with the additional criterion that we veto any candidate which passes the antiproton (proton) selection. The electron identification efficiency is 95\%, with pion and kaon fake rates $\thicksim 1 \%$.

We reconstruct candidates for both $\overset{\scriptscriptstyle{(-)}}{D^0} \rightarrow \bar{p} e^+$ and $\overset{\scriptscriptstyle{(-)}}{D^0} \rightarrow p e^-$ modes separately. We evaluate the difference between the beam energy and the sum of the electron and proton energies ($\Delta E$), and require $|\Delta E|$ to be within two standard deviations ($\sigma_{\Delta E} = 5.3$ MeV) of zero. For selected events, we compute the beam-constrained mass ~\cite{MarkIII}, defined as:
\be
\vspace{-5pt}
M_\text{bc}=\sqrt{E^2_\text{beam}-(\sum_i \mathbf{p}_i)^2},
\vspace{2pt}
\ee
where $E_\text{beam}$ is the beam energy and $\mathbf{p}_i$ represents the momenta of each final state particle. A signal would appear as a peak at the $D^0$ mass \cite{PDG}.
\section{Backgrounds and Signal Simulations}
\begin{figure}[ht]
\begin{center}
\centerline{ \epsfxsize=2.5in \epsffile{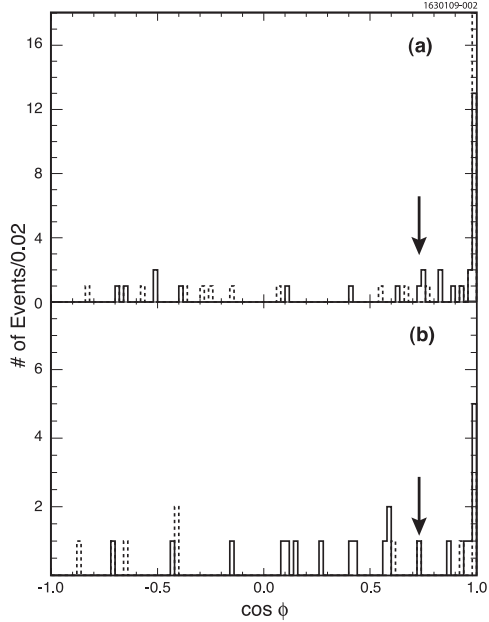}}
\caption{ \label{fig:costheta} Distributions of cos$\phi$, the angle between $e^+e^-$ candidates, as discussed in the text, for (a) continuum MC, and (b) data. The dotted histograms show cases where the $e^+$ is from a $\overset{\scriptscriptstyle{(-)}}{D^0}\rightarrow \bar{p} e^+$ candidate, and the solid histograms correspond to cases where the $e^-$ comes from a $\overset{\scriptscriptstyle{(-)}}{D^0}\rightarrow p e^-$ candidate. Events to the right of the arrows are eliminated.}
\end{center}
\end{figure}

Monte Carlo (MC) simulations are used to understand the response of the CLEO-c detector, to characterize and estimate the possible backgrounds, and to determine efficiencies of the reconstructed $D^0$ and $\bar{D}^0$ decay modes. In each case $e^+e^- \rightarrow \psi(3770) \rightarrow D \bar{D}$ events are generated with the $\mathrm{EVTGEN}$ program \cite{EvtGen}, and the response of the detector to the daughters of the $D\bar{D}$ decays is simulated with $\mathrm{GEANT}$ \cite{GEANT}. The $\mathrm{EVTGEN}$ program includes simulation of initial state radiation (ISR) events, i.e. events in which the $e^+$ and/or $e^-$ radiates a photon before the annihilation. The program $\mathrm{PHOTOS}$ \cite{PHOTOS} is used to simulate final state radiation (FSR). We use two types of MC events:
\begin{itemize}
\item continuum MC events, in which $e^+e^-$ annihilations into $\bar{u} u$, $\bar{d} d$ and $\bar{s} s$ quark pairs are simulated. It also includes the photon radiation by the initial state quarks.
\item signal MC events, in which either the $D^0$ or the $\bar{D^0}$ always decays in one of the two modes measured in this analysis while the other $\bar{D}^0$ or $D^0$, respectively, decays generically.
\end{itemize}
The decay of $D$ mesons into baryon pairs is kinematically forbidden, and so in the SM any real proton detected must be from a continuum event. Our largest source of potential background is the combination of a real proton from such an event with an electron from a photon pair conversion. We studied this background using a continuum MC simulation with five times the luminosity of our data sample. In Fig.~\ref{fig:costheta} we plot the $\cos\phi$ distribution, where $\phi$ is the angle between the $e^-$ and any other $e^+$ candidate. All selection requirements are applied, except that we relax the $\Delta E$ requirement to $\pm4 \sigma$, and accept candidates in the broader $M_\text{bc}$ range between $1.83$ and $1.89$ GeV. A clear excess near $\cos\phi=1$ is observed. We remove these events by requiring $\cos\phi<0.73$, which removes 71\% of the background with a 3.4\% loss in signal efficiency.

\begin{figure}[htb]
\begin{center}
\centerline{\epsfxsize=2.5in \epsffile{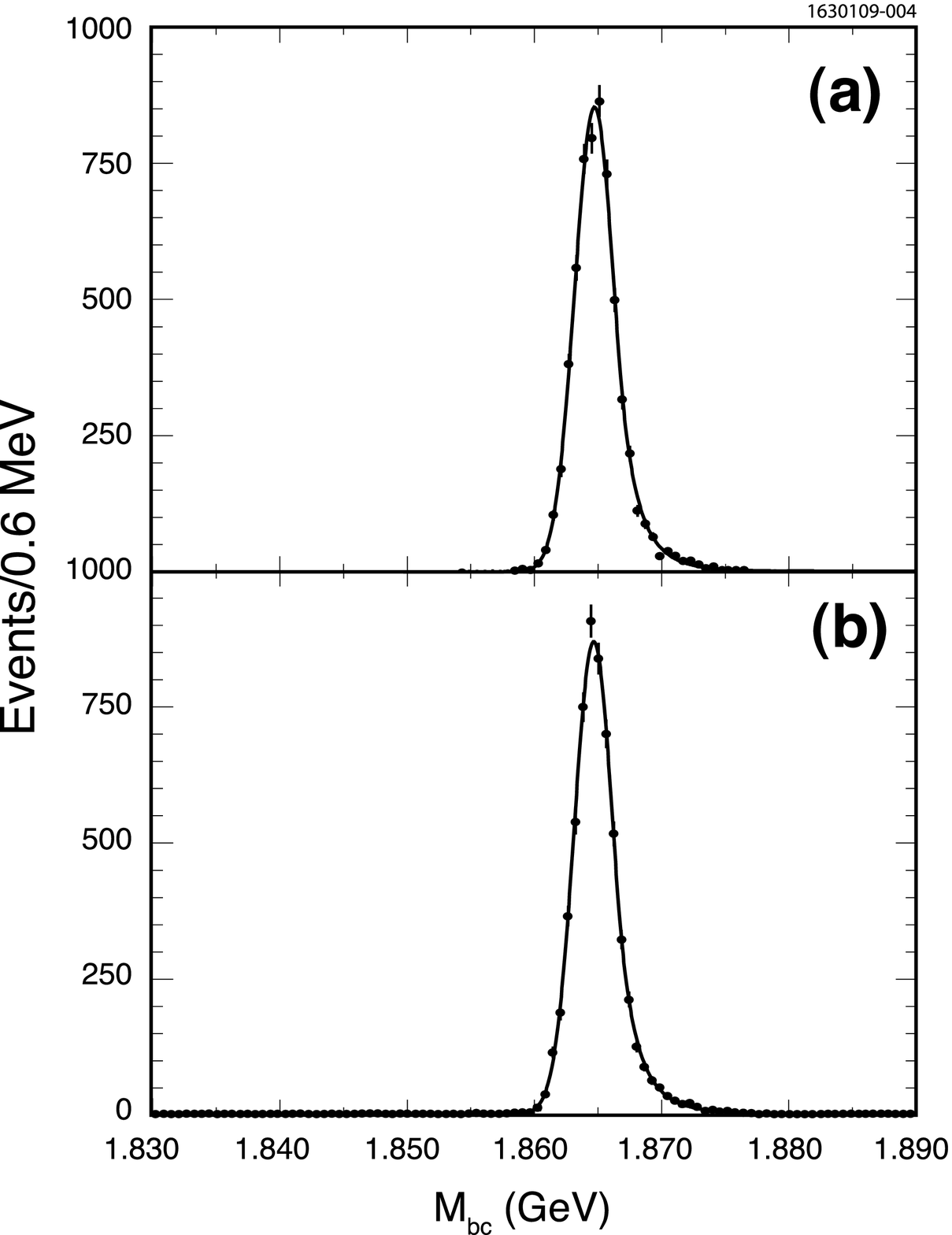}}
\caption{
\label{fig:MCmbc}$M_\text{bc}$ distributions for (a) $\overset{\scriptscriptstyle{(-)}}{D^0} \rightarrow \bar{p}e^+$, and (b)
for $\overset{\scriptscriptstyle{(-)}}{D^0} \rightarrow p e^-$ from signal MC, fitted with Crystal Ball functions.}
\end{center}
\vspace{1pt}
\end{figure}

We determine the $\overset{\scriptscriptstyle{(-)}}{D^0}$ signal line shape parameters and detection efficiencies using a signal MC sample for each mode. The $M_\text{bc}$ distributions are shown in Fig.~\ref{fig:MCmbc}.
We describe the signal shape using a Crystal Ball function \cite{CBL}, which has the form:
%\begin{widetext}
\[f(M_\text{bc}|M_\text{D},\sigma_{M_\text{bc}},\alpha,n)= \left( \begin{array}{l}
  A\;{\rm exp}\left[-{1\over 2}\left({{M_\text{bc}-M_\text{D}}\over \sigma_{M_\text{bc}}}
\right)^2\right]~~~~{{\rm for}~M_\text{bc}<M_\text{D}-\alpha\sigma_{M_\text{bc}}}\\
 A\;{{\left({n\over \alpha}\right)^n e^{-{1\over 2}\alpha^2}
\over \left({{M_\text{bc}-M_\text{D}}\over
\sigma_{M_\text{bc}}}+{n\over \alpha}-\alpha\right)^n}}
~~~~~~~~~~~{{\rm for}~M_\text{bc}>M_\text{D}-\alpha\sigma_{M_\text{{bc}}}}\\
{\rm here}~A^{-1}\equiv \sigma_{M_\text{bc}}\left[{n\over
\alpha}{1\over {n-1}}e^{-{1\over 2}\alpha^2} +\sqrt{\pi\over
2}\left(1+{\rm erf}\left({\alpha\over\sqrt{2}}\right) \right)\right],
\end{array}\right.\]
%\end{widetext}
\noindent where $A$ is an overall normalization, $M_\text{D}$ is the $D^0$ mass \cite{PDG}, $\sigma_{M_\text{bc}}$ is the mass resolution, and $n$ and $\alpha$ are parameters governing the shape of the high mass tail. This high mass tail results from initial state radiation from the $e^-$ and/or $e^+$ beams. In each fit, the parameters are determined by a binned maximum likelihood fit and their values are fixed in fits to data, with the exception of $n$. The fits are highly insensitive to the precise value of $n$, which is fixed to 7.0 throughout the analysis.

From the reconstructed yields, we determine signal efficiencies for $\overset{\scriptscriptstyle{(-)}}{D^0} \rightarrow \bar{p} e^+$ and $\overset{\scriptscriptstyle{(-)}}{D^0} \rightarrow p e^-$ of (59.1$\pm$0.5)\% and (59.4$\pm$0.5)\%, respectively.
\section{Results from Data}\label{result}
\begin{figure}[ht]
\begin{center}
\centerline{\epsfxsize=2.5in \epsffile{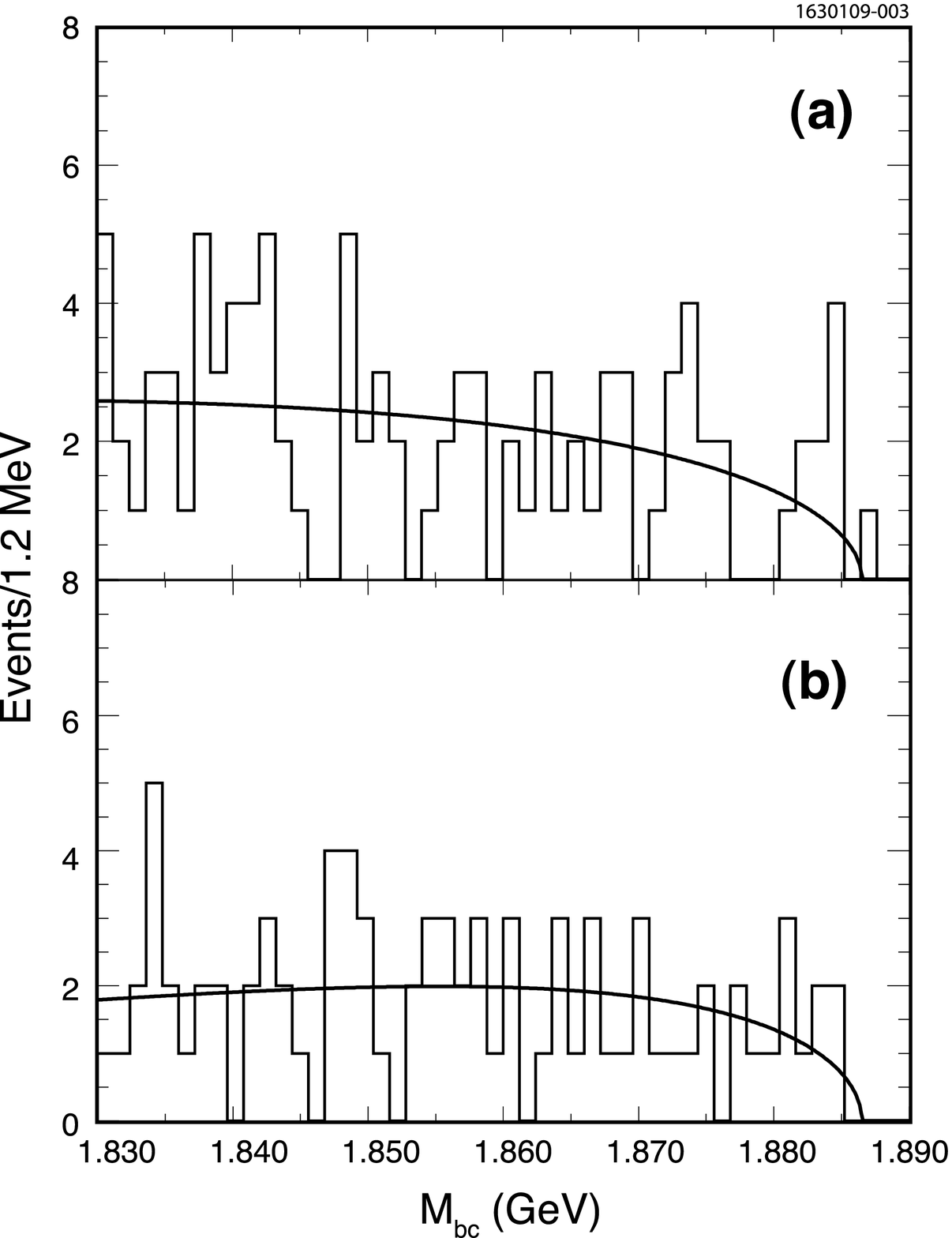}}
\caption{
\label{fig:bgdshape}{$M_\text{bc}$ distributions for (a) $\overset{\scriptscriptstyle{(-)}}{D^0} \rightarrow p e^-$ and (b) for $\overset{\scriptscriptstyle{(-)}}{D^0} \rightarrow \bar{p}e^+ $ from data shown by solid histograms. The curves are the fits as described in text.}}
\end{center}
\vspace{-20pt}
\end{figure}

The $M_\text{bc}$ distribution of events passing all selection criteria in data are shown in Fig.~\ref{fig:bgdshape} for $\overset{\scriptscriptstyle{(-)}}{D^0}\rightarrow \bar{p}e^+$ and $\overset{\scriptscriptstyle{(-)}}{D^0} \rightarrow p e^-$ modes separately. The background shape is parameterized by an ARGUS threshold function \cite{ARGUS}, which has the form:
\begin{equation}
 f(M_\text{bc})=KM_\text{bc}\sqrt{1 - \left({M_\text{bc} \over E_{beam}} \right)^2} {\rm exp} \left[{ S\left(1 -\left[ {M_\text{bc} \over E_{beam} }\right]^2\right)}\right]\label{Eq:Argus}
\end{equation}
Here, $K$ is an overall normalization, and the other parameters, $E_{beam}$ and $S$, govern the shape of the distribution; $E_{beam}$ is the beam energy and $S$ is a scale factor for the exponential. We fit the $M_\text{bc}$ distributions of the individual modes $\overset{\scriptscriptstyle{(-)}}{D^0} \rightarrow \bar{p}e^+$ and $\overset{\scriptscriptstyle{(-)}}{D^0} \rightarrow p e^-$ with fixed signal shape parameters (from the signal MC) and fix $E_{beam} = 1.8865$ GeV, but float parameters $K$ and $S$ in the background function. The fits are shown in Fig.~\ref{fig:bgdshape}(a) and Fig.~\ref{fig:bgdshape}(b). In both cases, the fit yield is zero and upper limits will be computed.
\section{Systematic Errors}\label{sec:errors}
We consider a number of systematic errors. We assign $\pm 0.7\%$ systematic error for finding each charged track, hence $\pm 1.4\%$ for both tracks. For electron identification we assign $\pm 1\%$ error \cite{elec}. The proton identification uncertainty is $\pm 1\%$, and was evaluated at higher beam energies on/near the Upsilon resonances by comparing the efficiency for identifying the proton in $\Lambda \rightarrow p  \pi^-$ decays in data and Monte Carlo simulation. Additional cross-checks were performed at center of mass energy, $E_{CM}=3770$ MeV that showed consistent performance of the particle identification over these running periods. Thus the overall particle identification uncertainty is $\pm 2\%$.

To estimate the systematic error arising from the $\Delta E$ cut, we compare signal yields using the nominal $\Delta E$ cut and a wide $\Delta E$ cut of $\pm$100 MeV for the kinematically similar $D^0 \rightarrow K^- \pi^+$ decay. The fractional decrease between the nominal and wide $\Delta E$ cuts are $9.02 \pm 0.34 \%$ and $8.93 \pm 0.14\%$ for data and MC simulation, respectively. The difference is $0.09 \pm 0.37 \%$ and therefore we assign a systematic uncertainty of $\pm 0.4\%$ to account for possible mismodelling of this quantity. The selection of cos$\phi<0.73$ reduces the efficiency by only 3.4\%, and we assign a $\pm 1\%$ uncertainty to the efficiency due to this cut. The uncertainties in the background shape due to the threshold $E_{beam}$ are determined by calculating the differences in the 90\% confidence level, (C.L) upper limit yields between the nominal fit and a fit with $E_{beam}$ shifted by $\pm 0.5$~MeV. Then the differences in the upper limits at the 90\% confidence level (C.L.) were taken to be the systematic errors from this source, which we estimate as $\pm$ 1\%. Uncertainties due to the signal shape parameters were found to be negligible. We also sought possible uncertainties due to differences in the veto efficiencies between data and simulation for kaons faking protons and antiprotons faking electrons and similarly for the charge conjugates. The differences were negligible. Finite MC statistics also introduces a $0.8\%$ systematic error. The systematic errors are summarized in Table~\ref{tab:Total}.
\begin{table}[htb]
\begin{center}
\caption{Systematic Uncertainties}
\begin{tabular}{lcr}
\hline \hline
Sources of errors & error($\pm\%$) \\ \hline
Tracking & 1.4 \\
Particle identification & $2$ \\
$\Delta E$ cut & 0.4  \\
cos$\phi$ cut & 1 \\
Background shape & 1 \\
Relative statistical error from signal MC & 0.8 \\
\hline
Total in quadrature & 3.0\\
\hline \hline
\end{tabular}
\label{tab:Total}
\end{center}
\end{table}
\section{Upper limits of Branching Fraction}\label{sec:upper}
\begin{figure}[htpb]
\begin{center}
\centerline{\epsfxsize=3.5in \epsffile{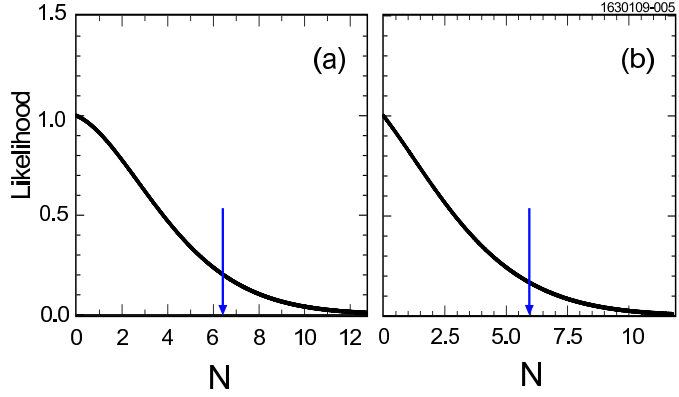}}
\caption{
\label{fig:likelihood}Fit Likelihood plots versus the yield $N$ for (a) for $\overset{\scriptscriptstyle{(-)}}{D^0}\rightarrow \bar{p}e^+ $ and (b) for $\overset{\scriptscriptstyle{(-)}}{D^0}\rightarrow p e^-$ from data. In each plot, the vertical line shows the value of $N$ below which 90\% of the total area lies.}
\end{center}
\vspace{-15pt}
\end{figure}

The likelihood distributions as a function of the assumed yields are shown in Fig.~\ref{fig:likelihood} for (a) the sum of possible $D^0 \rightarrow pe^-$ and $\bar{D}^0 \rightarrow pe^-$  yields, and for (b) the sum of possible $D^0 \rightarrow \bar{p}e^+$ and $\bar{D}^0 \rightarrow \bar{p}e^+$ yields. We determine the upper limits of branching ratios by integrating the likelihood function to include 90\% of the probability. We find 90\% confidence level (C.L.) upper limits of 6.40 and 6.00 events, respectively. We compute the upper limits on the branching fractions using:
\be \mathcal{B} = \frac{N}{\epsilon_{-} N_{D^0\bar{D}^0} }. \label{eq:BR} \ee
Here, $N_{D^0\bar{D}^0} = (1.031 \pm 0.008 \pm 0.013)\times 10^{6}$ is the number $D^0\bar{D}^0$ events at the $\psi(3770)$, where the first error is statistical and the second is due to systematics \cite{DNum}, $N$ is the 90\% C.L. upper limit and $\epsilon_{-}$ is the signal MC efficiency, reduced by one standard deviation. We determine an upper limit for the sum $\mathcal{B}(D^0 \rightarrow \bar{p} e^+) + \mathcal{B}(\bar{D}^0 \rightarrow \bar{p} e^+)$. We interpret this as a conservative upper limit on $\mathcal{B}(\bar{D}^0 \rightarrow \bar{p} e^+)$ or $\mathcal{B}(D^0 \rightarrow \bar{p} e^+)$. A similar interpretation is used for $\mathcal{B}(\overset{\scriptscriptstyle{(-)}}{D^0} \rightarrow p e^-)$. The calculated upper limits with and without the systematic errors are shown in Table~\ref{tab:result}. In particular, we find $\mathcal{B}(\overset{\scriptscriptstyle{(-)}}{D^0} \rightarrow \bar{p}e^+)<1.1 \times 10^{-5}$ and $\mathcal{B}(\overset{\scriptscriptstyle{(-)}}{D^0} \rightarrow p e^-)<1.0 \times 10^{-5}$, both at 90\% C.L.

\begin{table}[ht]
\begin{center}
\caption{Results from fits to the $M_\text{bc}$ distributions and the resulting upper limits on branching fractions for both of the modes.}
\begin{tabular}{ccc}
\hline \hline
  &$\overset{\scriptscriptstyle{(-)}}{D^0} \rightarrow \bar{p}e^+$ & $\overset{\scriptscriptstyle{(-)}}{D^0}\rightarrow p e^-$\\
  \hline
  Upper limit on N & 6.42 & 5.94 \\
  Upper limit on N(including systematic errors) & 6.61 & 6.12 \\
  Upper limit on $\mathcal{B}$ & $<1.1 \times 10^{-5}$ & $<1.0 \times 10^{-5}$\\
  \hline \hline
\end{tabular}
 \label{tab:result}
\end{center}
\end{table}
\section{Conclusions}
We have searched for the $B$ and $L$ violating decays $\overset{\scriptscriptstyle{(-)}}{D^0} \rightarrow \bar{p} e^+$ and $\overset{\scriptscriptstyle{(-)}}{D^0} \rightarrow pe^-$ and find no evidence of these decays. We obtain branching fraction upper limits of $\mathcal{B}(D^0\rightarrow \bar{p} e^+)[\mathcal{B}(\bar{D}^0 \rightarrow \bar{p} e^+)]<1.1 \times 10^{-5} $ and $\mathcal{B}(D^0\rightarrow p e^-)[\mathcal{B}(\bar{D}^0 \rightarrow p e^-)]<1.0 \times 10^{-5} $, both at 90\% C.L. Using these limits, and the $D^0$ lifetime, $\tau_{D^0} = (410.1 \pm 1.5) $ fs \cite{PDG}, we compute the partial widths ($\Gamma_i=\mathcal{B}_i/\tau_{D^0}$) to be:
\bea
\Gamma(\overset{\scriptscriptstyle{(-)}}{D^0} \to \bar{p} e^+)< 2.8\times 10^{7}~~{\rm s^{-1}} ~~~ {\rm and } && %\nonumber
\Gamma(\overset{\scriptscriptstyle{(-)}}{D^0} \to p e^-)< 2.5\times 10^{7}~~{\rm s^{-1}}.
\eea
These decay width limits provide less stringent constraints on new physics interactions than, for instance, proton decay experiments. However, no previous searches have investigated the possibility of charmed mesons violating $B$ and $L$. These limits do not violate the predictions of higher generation models, which predicts $\mathcal{B}(D^0 \rightarrow \bar{p} l^+) \sim 10^{-39}$ \cite{Nag}.

\section{Acknowledgments}
We gratefully acknowledge the effort of the CESR staff in providing us with excellent luminosity and running conditions. This work was supported by the National Science Foundation and the U.S. Department of Energy, the Natural Sciences and Engineering Research Council of Canada, and the U.K. Science and Technology Facilities Council.

\bibliographystyle{apsrev}
%\bibliography{}

\end{document}